\documentclass[oldversion]{aa}
\usepackage{txfonts}
\usepackage{natbib, graphicx}

\begin{document}
\title{The Galaxy Cross-Correlation Function as a Probe of the Spatial 
Distribution of Galactic Satellites}
\author{Jacqueline Chen \inst{1,2}}
\institute{Argelander-Institut f\"{u}r Astronomie, Universit\"{a}t Bonn,
Auf dem H\"{u}gel 71,
D-53121 Bonn; 
       {\tt  jchen@astro.uni-bonn.de}
       \and
Kavli Institute for Cosmological Physics, Dept. of Astronomy \& Astrophysics
University of Chicago,
5640 South Ellis Avenue,
Chicago, IL 60637; 
}

\abstract{The spatial distribution of satellite galaxies around host galaxies can illuminate the relationship between satellites and dark matter subhalos and aid in developing and testing galaxy formation models.  Previous efforts to constrain the distribution attempted to eliminate interlopers from the measured projected number density of satellites and found that the distribution is generally consistent with the expected dark matter halo profile of the parent hosts, with a best-fit power-law slope of $\approx -1.7$ between projected separations of $\sim30 h^{-1}$ kpc and $0.5 h^{-1}$ Mpc.  Here, I use the projected cross-correlation of bright and faint galaxies to analyze contributions from satellites and interlopers together, using a halo occupation distribution (HOD) analytic model for galaxy clustering.  This approach is tested on mock catalogs constructed from simulations.  I find that analysis of Sloan Digital Sky Survey (SDSS) data gives results generally consistent with interloper subtraction methods between the projected separations of $10 h^{-1}$ kpc and $6.3 h^{-1}$ Mpc, although the errors on the parameters that constrain the radial profile are large, and larger samples of data are required.}
\maketitle

\section{Introduction}
\label{sec:intro}

In the hierarchical assembly of dark matter (DM) halos, progenitor halos merge to form larger systems.  Some DM halos survive accretion and form a population of subhalos within the virial radius of the larger halo \citep[e.g.,][]{ghigna_etal98,klypin_etal99}.  Baryonic material cools and forms stars in the center of some halos and subhalos, resulting in galaxies and satellite galaxies.  Satellite galaxies are biased tracers of the potential well of their host DM halos, and satellite dynamics have been used to provide constraints on the total mass distribution in galactic halos \citep[e.g.,][]{zaritsky_white94,zaritsky_etal97,prada_etal03,vandenbosch_etal04,conroy_etal05,faltenbacher_diemand06} and to test the nature of gravity \citep{klypin_prada07}.  In addition, the spatial distribution of the satellite galaxy population reflects the evolution of dwarf galaxies and the mass accretion history of the halo.  For example, observations of satellites around early-type galaxies suggest that the distribution lies along the major axis of the light distribution \citep[e.g.,][]{sales_lambas04,brainerd05,yang_etal06,azzaro_etal07}, while in DM simulations the angular distribution of subhalos follows the shape of the DM halo, which is indicative of infall of subhalos along filaments \citep[e.g.,][]{zentner_etal05,libeskind_etal05}.  

The radial spatial bias of satellite galaxies is another important indicator, providing a simple observable for testing galaxy formation models.  Simulations of galaxy cluster-sized halos suggest that the distribution of DM subhalos has a radial profile that is less concentrated than the DM density distribution of the host halo at small radii (within $\sim  20-50\%$ of the virial radius) but has a profile similar to the DM halo at larger radii \citep{ghigna_etal98,colin_etal99,ghigna_etal00, springel_etal01,delucia_etal04,diemand_etal04,gao_etal04,nagai_kravtsov05,maccio_etal06}.  Simulations that include baryons, star formation and cooling, however, show that the distribution of galaxies associated with subhalos has a steeper inner profile than the subhalo distribution, both at cluster and at galaxy scales, and these distributions are in general agreement with observations \citep{nagai_kravtsov05,maccio_etal06}.  For samples of subhalos selected by dark matter mass, objects found near the halo center have lost a greater percentage of their dark matter mass due to tidal stripping than objects near the virial radius.  As  result, the concentration of the distribution of subhalos is lowered.  Stellar mass-selected samples of satellite galaxies are resistant to this effect since baryonic components are located in the centers of dark matter subhalos and are tightly bound.  

Observations of the Local Group dwarf population suggest a distribution more radially concentrated than that of DM subhalos in dissipationless numerical simulations \citep{kravtsov_etal04b,taylor_etal04,willman_etal04}.  At large radii, this distribution would also be inconsistent with the DM halo profile.  This strong radial bias, however, may not apply to brighter satellites -- such as the Magellanic Clouds -- and may be limited to the faintest dwarfs, related to the physics of the formation of the smallest dwarf galaxies \citep{kravtsov_etal04b,diemand_etal05} or incompleteness in observing faint objects at large radii.  

Constraints on the radial distribution of galactic satellites beyond the Local Group have been limited by the effect of interlopers -- projected objects contaminating the sample of satellite galaxies.  Previous attempts to constrain the satellite distribution have used galaxy redshift surveys and various methods of subtracting interlopers from samples of faint galaxies near bright, $\sim L_{*}$, galaxies.  \citet{vandenbosch_etal05} used data from the Two Degree Field Galaxy Redshift Survey (2dFGRS), but concluded that the data does not allow meaningful constraints on the radial distribution at small radii due
to the incompleteness of close pairs in the survey, although the data is consistent with the dark matter distribution at large radii. In an independent
analysis, \citet{sales_lambas05} fit a power-law slope to the
satellite distribution for projected
radii between 20 and 500 $h^{-1}$ kpc, finding a slope shallower than expected for the DM halo with a significant dependence on
morphological type of the parent galaxies but using no interloper subtraction.  \citet{chen_etal06} used data from the Sloan Digital Sky Survey (SDSS), testing methods of interloper subtraction on numerical simulations to show that the radial profile of satellites of isolated $\sim L_{*}$ galaxies is steeper than the subhalo profile in simulations and possibly consistent with being as steep as the DM profile.  

An alternate approach to this problem is to use the projected cross-correlation of bright and faint galaxies.  The correlation function is the statistical measure of the excess probability over a random distribution of finding pairs of objects at separation, $r$.  It can be calculated analytically from the mass function of halos, the spatial distribution of galaxies within halos, the halo bias, and the probability of finding a number of galaxies in a halo of mass, $M$, parameterized by the halo occupation distribution \citep[HOD;][]{ma_fry00,seljak00,scoccimarro_etal01,berlind_weinberg02}.  Many recent works have used such analyses of autocorrelated samples of galaxies in the SDSS in order to constrain the relation between distributions of galaxies and dark matter and to constrain cosmological parameters \citep[e.g.,][]{zehavi_etal02,zehavi_etal04,zehavi_etal05,abazajian_etal05,zheng_etal07,tinker_etal08}. 

A halo model analysis of the clustering of  galaxies can constrain both the satellite and interloper populations, without the need to cut the bright galaxy sample by environment or use subtraction methods in order to eliminate interlopers.   Here, I use the cross-correlation of bright and faint galaxies, as well as the autocorrelation functions of bright and faint galaxies, to constrain the spatial distribution of faint satellites around bright galaxies, using data from the SDSS spectroscopic sample.  The ability of the data to constrain the spatial distribution of satellites with this method is tested with mock data sets from simulations populated with galaxies by a realistic HOD.  Best-fit parameters and errors are characterized using Monte Carlo Markov chains (MCMC).  Despite the degeneracies between parameters, constraints on the spatial distribution appear without systematic bias.  Errors on the parameter estimates, however, are large in the current data set, and larger datasets are required. 

The paper is organized as follows.  The SDSS data used in the analysis are discussed in Section \ref{sec:sdss}, while the correlation function and associated statistical and systematic errors are discussed in Section \ref{sec:corr}.  The HOD analytic model is detailed in Section \ref{sec:hm}, and modeling with MCMC and populated simulations is discussed in Section \ref{sec:mcmc}.  The main results are presented in Section \ref{sec:results}. Conclusions
are summarized in Section \ref{sec:conclusions}.  Throughout, I assume flat $\Lambda$CDM cosmology with $\Omega_{\rm m}=0.3$, $\Omega_{\rm b}=0.04$, $\sigma_{8}=0.9$, 
and $h=0.7$.

\section{The Sloan Digital Sky Survey}
\label{sec:sdss}

The Sloan Digital Sky Survey (SDSS) will image up
to $10^4$ deg$^2$ of the northern Galactic cap in five bands,
$u,g,r,i,z$, down to $r \sim 22.5$ and acquire spectra for galaxies with $r$-band Petrosian magnitudes
$r \leq 17.77$ and $r$-band Petrosian half-light surface brightnesses
$\mu_{50} \leq 24.5$ mag arcsec$^{-2}$, using a
dedicated 2.5m telescope at Apache Point Observatory in New Mexico \citep{fukugita_etal96,gunn_etal98,york_etal00,hogg_etal01,smith_etal02,strauss_etal02,blanton_etal03b,gunn_etal06,tucker_etal06}.  An automated pipeline measures the
redshifts and classifies the reduced spectra \citep[D. J. Schlegel et
al. 2009, in preparation]{stoughton_etal02,pier_etal03,ivezic_etal04}.

For this analysis, I use a subset of the spectroscopic main
galaxy catalog available as Data Release Five (DR5) and including all of the data available as of Data Release Four (DR4) \citep{adelman_mccarthy_etal06}.  I refer to this sample hereafter as DR4+.  Because the SDSS spectroscopy is taken through
circular plates with a finite number of fibers of finite angular size,
the spectroscopic completeness varies across the survey area. The
resulting spectroscopic mask is represented by a combination of disks
and spherical polygons \citep{tegmark_etal04}.  Each polygon also
contains the completeness, a number between 0 and 1 based on the
fraction of targeted galaxies in that region which were observed. I
apply this mask to the spectroscopy and include only galaxies from
regions where completeness is at least 90\%, for a completeness-weighted area of 5104 ${\rm deg^2}$.  I use $r$-band magnitudes in DR4+, built from the NYU Value-Added Galaxy Catalog \citep{blanton_etal05}, normalized to $h$=1, such that 
\begin{equation}
M_{r} \equiv M_{0.1_{r}} - 5{\rm log}_{10} h,
\end{equation}where $M_{0.1_{r}}$ is the absolute
magnitude K-corrected to $z$=0.1 ({\tt kcorrect v3.4}) as described in
\citet{blanton_etal03c}.  

A volume limited sample of galaxies down to $M_{r} = -18$ contains galaxies out to redshift $z=0.048$, with a  median galaxy redshift of z=0.038.  The redshift limit is a trade-off with the magnitude limit.  I create a `bright' sample and a `faint' sample, defined such that bright galaxies have r-band magnitudes between $-21$ and $-20$ and faint galaxies have magnitudes between $-19$ and $-18$.  The bright galaxy sample contains galaxies that are near $L^{*}$, while the faint galaxy sample contains galaxies that have luminosities similar to bright satellites such as the Magellanic Clouds.  By excluding objects between $M_{r}= -19$ and $M_{r}= -20$, the samples are fairly analogous to the samples used in previous works that constrained the spatial distribution of satellite galaxies \citep[e.g.,][]{chen_etal06}.  The corresponding number density of objects for these samples is calculated from the data and shown in Table \ref{tab:numdens}, along with the total number of objects in each sample.

In addition to the faint and bright samples, a luminosity-threshold sample of $M_{r} < -21$ and a luminosity-threshold sample of $M_{r} < -19$ are created from the same volume-limited sample.  These samples are necessary in order to constrain the masses of the halos in which bright and faint galaxies are found, as further described in Section \ref{sec:hm}.

\begin{table}[h]
\caption{Number Densities of Volume-Limited Samples of SDSS Galaxies}
\label{tab:numdens}
\begin{tabular}{lll}
\hline
$M_{r}$ & $\bar{n}$ ($h^{3} {\rm Mpc}^{-3}$) &N\\
\hline\hline
$-21$ to $-20$ (`bright') & 0.00392 & 5863 \\
$-19$ to $-18$ (`faint')  & 0.01137 & 16992\\
$< -19$ & 0.012549 & 19085\\
$< -21$ & 0.00056 & 846\\
\hline
\end{tabular}
\end{table}

\section{Estimating the Correlation Function}
\label{sec:corr}

The correlation function measures the excess probability of finding a pair of objects at some separation, comparing counts of pairs of objects in catalogs of real objects (DD) and in catalogs of objects with randomized positions (RR) and pairs where one object is in the real catalog and its counterpart in the random catalog (DR or RD).\footnote{For the SDSS samples, the random catalogs contain at least twenty times the number of objects in the real catalogs.  In order to create a random catalog from a data catalog, the objects in the data catalog are sampled with replacement -- such that each data point may be sampled several times -- and then assigned a randomized position on the sky.}  The total number of pairs is normalized by the total number of objects in a sample, such that 
\begin{equation}
DD = \frac{\rm number~ of ~pairs}{N_{D}~N_{D}}.  
\end{equation}
Using the \citet{landy_szalay93} estimator, the cross-correlation as a function of projected separation, $r_p$, and line-of-sight separation, $\pi$, is
\begin{equation}
\xi(r_p,\pi)=\frac{DD-DR-RD+RR}{RR},
\end{equation}
while the autocorrelation is
\begin{equation}
\xi(r_p,\pi)=\frac{DD-2DR+RR}{RR}.
\end{equation}

The parallel and perpendicular components of the pair separations are distinguished in the data by 
\begin{equation}
\pi \equiv \frac{{\bf s \cdot l}}{\mid {\bf l } \mid}
\end{equation}
and
\begin{equation}
r_p^2  \equiv {\bf s \cdot s } - \pi^2,
\end{equation}
where ${\bf s} \equiv {\bf v_1} - {\bf v_2}$,  ${\bf l} \equiv \frac{1}{2} ({\bf v_1} + {\bf v_2})$ and ${\bf v_1}$ and ${\bf v_2}$ are the redshift positions of the two objects in a pair \citep{fisher_etal94}.  

From the correlation function, the {\it projected} correlation function is 
\begin{equation}
w_p(r_p) = 2 \int \xi(r_p,\pi) d\pi.
\end{equation}
where $\pi$ is integrated to $\pi = 40 h^{-1}$ Mpc, large enough to include all significant clustering signal.  The projected correlation is measured in logarithmic bins for projected radii between 10 $h^{-1}$ kpc to 6.3 $h^{-1}$ Mpc, with 14 total data points.  

\begin{figure}
\centering
\resizebox{3in}{!}
	{\includegraphics{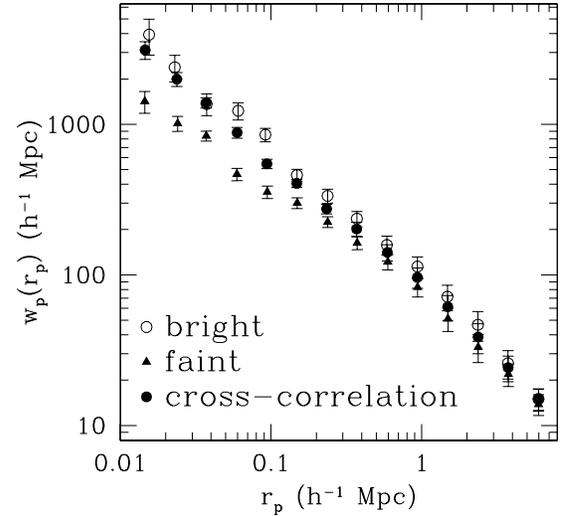}}
\caption{The measured projected correlation functions in the SDSS data, where the cross-correlation of bright and faint galaxies is indicated by filled circles, the bright autocorrelation by open circles, and the faint autocorrelation by filled triangles.  Bright galaxies have $r$-band magnitudes between $-21$ and $-20$, while faint galaxies have $r$-band magnitudes between $-19$ and $-18$.}
\label{fig:data}
\end{figure}

Statistical errors in the correlation are estimated via jackknife sampling \citep[see, e.g.,][]{lupton93}.  In the SDSS data, the survey area is divided into 205 equal area samples using the hierarchical pixel scheme SDSSPix\footnote{See http://lahmu.phyast.pitt.edu/$\sim$scranton/SDSSPix.}, which represents well the rectangular geometry of the SDSS stripes.  The covariance matrices are estimated by the covariance between jackknife samples -- samples where one of the 205 equal area subsamples is omitted.  In order to appropriately estimate the covariance matrix, the number of jackknife samples must be significantly greater than the number of bins in which the correlation function is estimated \citep{hartlap_etal07}.  Typically, each jackknife sample is constructed to be larger than the largest separation measured in the corresponding correlation function, and the number of jackknife samples is at least equal to the square of the number of bins.  With 14 bins, 196 is the minimum number of jackknifes samples.  The data sample uses slightly more, 205.  The typical jackknife region, then, has a comoving volume of $\sim19^3 h^{3} \rm Mpc^{-3}$.  The projected cross-correlation and corresponding autocorrelation functions are shown in Fig. \ref{fig:data}.

In the SDSS spectroscopic survey, no single pointing of the telescope can measure the spectra of objects that are separated by less than $55\arcsec$, the fiber collision distance.  At the median redshift of the sample, $55\arcsec$ is 30 $h^{-1}$ kpc, well above the minimum separation of 10 $h^{-1}$ kpc in the data.  To achieve the minimum separation, I employ a catalog corrected for collisions.  For pairs of objects separated by less than the fiber collision distance in the target selection catalog, the unobserved target is assigned the redshift of the observed object in the pair.  This could bias the sample by overestimating pairs, but significant error is unlikely given that pairs at those separations are extremely correlated (as in Fig. \ref{fig:data}).

Previous studies have described systematic errors with angular dependence \citep{masjedi_etal06,mandelbaum_etal06b,adelman_mccarthy_etal08}.  Issues with sky subtraction may lead to underestimation of pairs by $\sim5\%$ with angular separations between $40-90\arcsec$ and an overestimation of pairs at angular separations less than $20-30\arcsec$.  This effect is most acute for pairs of objects that consist of an apparently bright object and an apparently faint object (Rachel Mandelbaum 2006, private communication).

In the cross-correlated sample, the estimated systematic error of 5\% for pairs with angular separations between $40\arcsec$ and $90\arcsec$ is smaller than the statistical errors in the correlation function, $\sim10\%$, indicating that such errors do not affect the results.  For the unconstrained overestimate of pairs with angular separations less than $20\arcsec$, only the first bin of the estimated cross-correlation (containing objects with projected physical separation of less than 16 $h^{-1}$ kpc) is affected.  In this bin, 23\% of the pairs could be possibly affected, while the fractional statistical error in the correlation function is 15\%.  Given that only one of the data points is affected, it seems unlikely that the parameter estimates are biased by this systematic.  However, I test the possible effects of this error in Sections \ref{sec:systematictest} and  \ref{sec:robustnessresults}.

\section{Halo Model of Correlation Functions}
\label{sec:hm}

The correlation function can be modeled using the halo model -- assuming that all galaxies and dark matter particles live in dark matter halos.  First, the correlation function is decomposed into two parts:  a one-halo term and a two-halo term, 
\begin{equation}
\xi(r)=\xi_{\rm 1h}(r) + \xi_{\rm 2h}(r),
\end{equation}  
where the terms correspond to 
contributions from galaxy pairs found in the same dark matter halo and 
galaxy pairs found in different halos.  At small separations, $r$, the one-halo term dominates the correlation function, while at large separations, the two-halo term dominates.  

All galaxies are identified as either a central galaxy, found at the center of a dark matter halo, or a satellite galaxy, found elsewhere in the halo.  Following \citet[][see their Appendix A \& B]{tinker_etal05}, the one-halo term is decomposed into two additional terms:  (1) a central-satellite term where one galaxy of a pair is the central
galaxy and the other is a satellite of the central galaxy and (2) a
satellite-satellite term where both galaxies in a pair are satellite
galaxies in a halo.  As in \citet[Eqn. 11]{berlind_weinberg02}, the one-halo term can be written, 
\begin{eqnarray}
 \xi_{\rm 1h}(r)  =   \frac{1}{2 \pi r^2 \bar{n}_{1}\bar{n}_{2}} &
 \times  \int_0^{\infty} dM \frac{dn}{dM} \langle N_{\rm pair}\rangle_{M}
\frac{1}{2R_{\rm vir}(M)} F^{\prime} \left(\frac{r}{2R_{\rm vir}} \right),
\end{eqnarray}
where $\bar{n}_1$ and $\bar{n}_2$ are the number densities of galaxies, and here and throughout this section `1' and `2' are used to denote generic samples.  $dn/dM$ is the mass function of halos, and $\langle N_{\rm pair} \rangle_{M}$ is the average number of pairs of objects in a halo of mass, $M$.  $R_{\rm vir}$ is the virial radius of a halo of mass, $M$, and $F(x)$ is the average fraction of galaxy pairs in such a halo with separation less than $r$, and $F^{\prime}(x)$ is the derivative with respect to $r$.  The virial radius of a halo is defined as the radius at which the mean mass density enclosed is 200 times the mean matter density of the universe. 

To separate the central-satellite term from the satellite-satellite term, for the cross-correlation function,
\begin{eqnarray}
\langle N_{\rm pair}\rangle_{M} F^{\prime} \left(x \right) = & ( \langle N_{\rm cen,b}N_{\rm sat,f} \rangle + \langle N_{\rm cen,f}N_{\rm sat,b} \rangle) F^{\prime}_{\rm cen-sat} (x) \nonumber \\
 &  + 
 \langle N_{\rm sat,b}N_{\rm sat,f} \rangle F^{\prime}_{\rm sat-sat} (x)
\end{eqnarray}
and for the autocorrelation,
\begin{eqnarray}
\langle N_{\rm pair}\rangle_{M} F^{\prime} \left(x \right) = &\\
& \langle N_{\rm cen}N_{\rm sat} \rangle F^{\prime}_{\rm cen-sat} (x)  + 
\frac{ \langle N_{\rm sat}(N_{\rm sat}-1) \rangle}{2} F^{\prime}_{\rm sat-sat} (x), \nonumber
\end{eqnarray}
where centrals are denoted by `cen,' satellites are denoted by `sat' and bright and faint objects are labeled `b' and `f.'  $F^{\prime}_{\rm cen-sat}$ is proportional to the volume-weighted halo density profile and $F^{\prime}_{\rm sat-sat}$ is proportional to the convolution of the halo density profile with itself.  Density profiles are modeled with NFW profiles \citep{navarro_etal97} and the convolution is calculated analytically as derived by \citet{sheth_etal01} and \citet{zheng04}.  The concentration of the dark matter halos is calculated following \citet{bullock_etal01} from a simple expression for concentration as a function of the redshift of collapse for a halo.  The galaxy distribution within the dark matter halos is parameterized by a scaling factor, such that $c_{\rm gal} = f c_{\rm dm}$.  Finally, the mass function of halos from \citet{jenkins_etal01} is employed.  This fitting formula approximates the mass function of halos with an accuracy better than $\approx 20\%$ down to halos of $10^{10} h^{-1} M_{\sun}$, which is well below the range of masses used in the data and populated simulations (see Sections \ref{sec:mcmc} and \ref{sec:results}).  As mentioned previously, a flat $\Lambda$CDM cosmology with $\Omega_{\rm m}=0.3$, $\Omega_{\rm b}=0.04$, $\sigma_{8}=0.9$, 
and $h=0.7$ is used.  

The two-halo term is more easily calculated in Fourier space, where the power spectrum $P(k)$ is the Fourier transform of the correlation function $\xi(r)$:
\begin{displaymath}
P_{\rm 2h}(k,r) = 
\end{displaymath}
\begin{displaymath}P_{m}(k) \left[\frac{1}{\bar{n}_1} \int dM_1 \frac{dn}{dM_1} \langle N \rangle_{M_1} b_h(M_1,r) y_g(k,M_1)\right] 
\end{displaymath}
\begin{equation}
~~~~~~~~~~~ \times \left[\frac{1}{\bar{n}_2} \int dM_2 \frac{dn}{dM_2} \langle N \rangle_{M_2} b_h(M_2,r) y_g(k,M_2)\right],
\end{equation}
where $y_g(k,M)$ is the Fourier transform of the halo density profile and $\langle N \rangle_{M}$ is the average number of objects in a halo of mass, $M$.  Common simplifications to this calculation have employed the large-scale bias, $b(M)$, and approximated the matter power spectrum, $P_{m}(k)$, with the linear power spectrum, $P_{\rm lin}(k)$. Here, I use the dark matter power spectrum of \citet{smith_etal03} and the scale-dependent halo bias such that 
\begin{equation}
b^2(M,r) = b^2(M) \frac{\left[1 + 1.17 \xi_{m}(r)\right]^{1.49}}{\left[1 + 0.69 \xi_{m}(r)\right]^{2.09}},
\end{equation}
where the scale-independent bias is provided by \citet[][see their Appendix A]{tinker_etal05}.  They  calibrate the analytic model of \citet{sheth_tormen99} with numerical simulations.  In addition, the assumption that the separation between halos is much greater than the scale of the halos leads to an overestimate in power at intermediate separations where the one-halo and two-halo terms are comparable -- counting pairs in overlapping halos that would rightly be counted as single halos.  I use the spherical exclusion approximation from \citet{tinker_etal05}.  They only specify the two-halo term for the autocorrelation.  For the cross-correlation, the geometric mean of the the two autocorrelated samples is used as an approximation, which performed well in tests on populated simulations.

While the terms in the calculation for the halo mass function, halo bias, and halo density profile are fixed, the number of galaxies and their spatial distribution in halos is parameterized and fit to the data.  I use the HOD prescription for characterizing the galaxy bias, which has been shown in simulations and semi-analytic models to be reasonable \citep{berlind_etal03,zheng_etal05}.  For luminosity-threshold samples (samples that include all objects brighter than a threshold luminosity) the average number of central galaxies per halo is a step function at a minimum mass, $\langle N_{\rm cen} \rangle = 1 $ for $M > M_{\rm min}$ and $\langle N_{\rm cen} \rangle  = 0$, otherwise.  For luminosity-binned samples (including all objects within a minimum and maximum luminosity), $\langle N_{\rm cen} \rangle$ is a step function, with both a minimum mass and a maximum mass:  $\langle N_{\rm cen} \rangle = 1 $ for $M_{\rm min} < M < M_{\rm max (cen)}$.\footnote{\citet{tinker_etal07} suggest a more physical model for $\langle N_{\rm cen}\rangle$ that takes into account the scatter between mass and 
luminosity.  Tests comparing the simple model for  $\langle N_{\rm cen}\rangle$ used in this work to their model, however, find no significant effect on the resulting correlation function.}

For the average number of satellite galaxies, a simple functional form with three parameters is used \citep{tinker_etal05}:
\begin{equation}
\langle N_{\rm sat} \rangle = \left( \frac{M}{M_{1}} \right)^{\alpha} {\rm exp}\left(-\frac{M_{\rm cut}}{M}\right).
\end{equation} 
Using luminosity-threshold samples, simulations show that $\langle N_{\rm sat} \rangle$ should ``roll-off'' faster than a power-law for halo masses near the minimum mass, and HOD parameterizations that account for ``roll-off'' provide good fits to observations \citep{zehavi_etal05}.  In addition, \citet{conroy_etal06} suggests that by adding a parameter for ``roll-off,'' best-fits for the slope $\alpha$ may be driven to $\alpha=1$.  However, for luminosity-binned samples -- as the cross-correlation uses --  parameterization of ``roll-off'' is not necessarily the same as is expected in luminosity-threshold samples. 

For each galaxy sample, there are three free HOD parameters  -- $M_{1}$, $\alpha$, and $M_{\rm cut}$ --  and two fixed HOD parameters -- $M_{\rm min}$ and $M_{\rm max(cen)}$.  $M_{\rm min}$ is set by the observed number density of objects in the faint or bright sample such that 
\begin{eqnarray}
n_{\rm gal ~(f/b)}  & =  & \int_{M_{\rm min ~(f/b)}}^{\infty}  (\langle N_{\rm cen ~(f/b)} \rangle + \langle N_{\rm sat ~(f/b)} \rangle) \frac{dn}{dM} dM\nonumber \\
& =  & \int_{M_{\rm min ~(f/b)}}^{M_{\rm max(cen) ~(f/b)}} \frac{dn}{dM} dM \nonumber \\
&&~~~~~ + \int_{M_{\rm min ~(f/b)}}^{\infty}  \langle N_{\rm sat ~(f/b)} \rangle \frac{dn}{dM} dM
\end{eqnarray} where ${\rm (f/b)}$ is used to note that the equation must be solved twice, once for the faint sample and once for the bright sample.  $M_{\rm max(cen)}$, however, cannot be set using those samples.  Instead, it must be set by the number density of objects in samples that contains objects brighter than the objects in the corresponding bright or faint sample such that
\begin{equation}
n_{\rm gal ~(>f/>b)} = \int_{M_{\rm max(cen) ~(f/b)}}^{\infty} (\langle N_{\rm cen ~(>f/>b)} \rangle + \langle N_{\rm sat ~(>f/>b)} \rangle) \frac{dn}{dM} dM,
\end{equation} where ${\rm (>f/>b)}$ is used to denote samples that contain objects brighter than the faint or bright sample.

In addition to the HOD parameters described above, parameters to describe the spatial distribution of objects are required.  As mentioned previously, the spatial distribution of galaxies is parameterized by $f$, a scaling factor relating the concentration of a DM halo to the concentration for the distribution of galaxies in that halo, such that $c_{\rm gal} = f c_{\rm dm}$.  The three clustering measures -- the faint autocorrelation, the bright autocorrelation, and the cross-correlation -- require a total of two such parameters.  The first, $f$, parameterizes the spatial distribution of the autocorrelated samples -- i.e., the distribution of any-sized satellite in any mass halo.  This comprises a large catalog of objects and smears out environmental differences, e.g., assuming that $f_{\rm bright} = f_{\rm faint}$.  The second, $f_{\rm cross}$ parameterizes {\it only} the distribution of faint satellite galaxies in halos with bright central galaxies, as defined in the samples.  This comprises a small number of objects and ensures that the parameter in question measures only the spatial distribution of satellite galaxies in galactic halos with $\sim L^*$ galaxies.  The narrowness of the category for $f_{\rm cross}$, however, means that the parameter is constrained by fewer data points than $f$ and depends only on the one-halo central-satellite term of the cross-correlation function, a term which is only dominant in the correlation function for separations less than $\sim$0.1 $h^{-1}$ Mpc.   

I simultaneously fit the cross-correlation function of the bright and faint samples, the autocorrelation function of the bright sample, and autocorrelation of the faint sample, which requires eight free HOD parameters: 3 parameters -- $M_1$, $\alpha$, and $M_{\rm cut}$ -- for the faint and bright sample each and 2 parameters, $f$ and $f_{\rm cross}$,  to describe the spatial distribution of objects in halos.

\section{Parameter Fitting with Monte Carlo Markov Chains}
\label{sec:mcmc}

\subsection{MCMC}

To estimate the best-fit parameters and constrain the errors in
their values, I use a Monte Carlo Markov Chain (MCMC)
method.  The algorithm explores the probability distributions of a
large number of parameters such that the distribution of points in parameter space visited by an infinitely long MCMC chain follows the underlying
probability density exactly.  MCMC 
both measures the best-fit parameters and constrains the errors in the 
parameter estimation and a sufficiently long chain will not be hampered 
by local minima in the parameter space.  

A MCMC chain is created by choosing points in the parameter space based on a trial distribution around the previous point in the chain, stepping along
the eigenvectors of the covariance matrix of the parameters.  Trial steps that result in a smaller $\chi^2$ are
accepted while steps that do not are accepted with some probability, $P = e^{-\Delta \chi^{2}/2}$.  
For a finite chain, when the distribution of steps reflects the underlying distribution with sufficient accuracy, the chain is said to have converged.  In addition, if a chain is started far outside the region of high probability, an initial section will be unrepresentative and must be discarded;  this initial section is referred to as `burn in.'  This truncation is unnecessary or can be minimized if the chain is started from a point already known to lie in the region of high probability.  For my chains, I start the MCMC at the point in parameter space that has been identified as the likelihood maximum using a simple downhill simplex $\chi^2$ minimization algorithm for the data.  For good measure, the trial steps are examined for large-scale modes indicative of burn-in and convergence is tested by calculation of the power spectrum of the chain as is described by \citet{dunkley_etal05}.

\subsection{Testing Parameter Estimates with  Mock Galaxy Catalogs}

To test the constraining power of my sample using this method, I create test catalogs, populating halos from a numerical simulation with reasonably-chosen values for the HOD parameters.\footnote{In this case, values are chosen to correspond roughly to the best-fit parameter values found for the observed sample.}  Analyzing this mock data and comparing the derived HOD parameters to the input will show how well the analysis can work and how good the data needs to be for the analysis to provide robust constraints.  The HOD parameters are listed in Table \ref{tab:popsims}.  For the spatial distribution, both $f=1$ and $f_{\rm cross}=1$.  MCMC is used to find the best-fit values of the eight free parameters and characterize the error distributions. 

\begin{table}[h]
\caption{HOD Parameters for Mock Data Sets}
\label{tab:popsims}
\begin{tabular}{llllll}
\hline
Catalog &
$M_{\rm min}$ &
$M_{\rm max(cen)}$ &
$M_{1}$ &
$\alpha$ &
$M_{\rm cut}$ \\
\hline\hline
bright & $1.4 \times 10^{12}$ & $9.9 \times 10^{12}$ & $2.0 \times 10^{13}$ & 1.0 &  $2.0\times 10^{12}$  \\
faint & $2.5 \times 10^{11}$ & $4.9 \times 10^{11}$ & $6.7 \times 10^{12}$ & 0.9 &  $3.5 \times 10^{11}$  \\
\hline
\end{tabular}

{\small Note -- All masses are in $h^{-1} M_{\sun}$.}
\end{table}

The simulation used here was performed using the Hashed Oct-Tree code of \citet{warren_salmon93}, and is similar to those presented in \citet{seljak_warren04} and \citet{warren_etal06}.   The simulation box is $400 h^{-1}$ Mpc on a side with $1280^3$ particles and a particle mass of $2.5 \times 10^{9} h^{-1} M_{\sun}$.  Halos are identified by the friends-of-friend technique \citep{davis_etal85} with a linking length of 0.2 times the mean interparticle separation.  For galaxies in my faint sample, the corresponding halo in the simulation has roughly $\sim$100 particles.  The simulation uses a cosmology of $\Omega_{m} + \Omega_{\Lambda} =1$, $\Omega_m = 0.3$, $\sigma_8=0.9$, $h=0.7$, and $\Omega_b=0.04$;  I use the output at the redshift of $z=0.0$.  

I create two volumes of mock data.  In the first, I use the full simulation box to look for biases in finding the best-fit values for the HOD parameters and to describe the degeneracies between parameters.  In the second, I use a volume of data similar to the volume of available data to show how well the analysis will work on the real data and how robust the the constraints will be.  I also use this volume of data to test systematics in the SDSS data and the ability of the method to recover different values for $f$ and $f_{\rm cross}$.

\subsection{Tests Using Idealized Mock Data Set}

The full simulation box of 400 $h^{-1}$ Mpc on a side is 42 times the volume of the volume-limited SDSS data set.  For this box, I calculate the bright and faint galaxy autocorrelations, the cross-correlation, and all associated covariance matrices using 400 jackknife samples.  Using the full simulation box, the diagonal errors on the faint autocorrelation and the cross-correlation functions are significantly less than 5\%.  The errors on the dark matter power spectrum \citep{smith_etal03}, however, are likely larger than 5\%.  Analyzing the full simulation box, then, will only constrain the errors on the dark matter power spectrum and say nothing about the biases in and degeneracies between HOD parameters.  In order to evade this problem, I replace the measured faint autocorrelation and the cross-correlation with the exact values for the chosen HOD.

\begin{table}[h]
\caption{Unmarginalized, Best-Fit Parameters for Mock Data Sets}
\label{tab:popsims_results}
\begin{tabular}{llll}
\hline
\multicolumn{1}{c}{Parameter} &
\multicolumn{1}{c}{Input Value} &
\multicolumn{2}{c}{Best-Fit with 68\% errors}\\
\multicolumn{1}{c}{}&
\multicolumn{1}{c}{}&
\multicolumn{1}{c}{Idealized} &
\multicolumn{1}{c}{115 $h^{-1}$ Mpc box } \\
\hline\hline
faint: & & &\\
~~~~~~~~~~log($M_{1}$) & 12.826 & $12.836^{+0.032}_{-0.042}$ & $12.673^{+0.195}_{-0.149}$ \\
&&& \\
~~~~~~~~~~$\alpha$      & 0.9    & $0.900^{+0.020}_{-0.024}$ & $0.853^{+0.124}_{-0.094}$\\
&&&\\
~~~~~~~~~~log($M_{\rm cut}$) & 11.544 & $11.561^{+0.057}_{-0.082}$ & $11.691^{+0.316}_{-0.378}$\\
&&&\\
\hline
bright: & &&\\
~~~~~~~~~~log($M_{1}$) & 13.301 & $13.341^{+0.041}_{-0.079}$ & $13.304^{+0.131}_{-0.275}$ \\
&&&\\
~~~~~~~~~~$\alpha$      & 1.0    & $1.042^{+0.041}_{-0.071}$ & $1.055^{+0.122}_{-0.238}$\\
&&&\\
~~~~~~~~~~log($M_{\rm cut}$)& 12.301 & $12.157^{+0.223}_{-0.218} $ & $10.888^{+1.654}_{-1.888}$\\
&&&\\
\hline
&&& \\
$f$                & 1.0    & $1.017^{+0.119}_{-0.159} $ & $0.890^{+0.809}_{-0.446}$\\
&&&\\
$f_{\rm cross}$    & 1.0    & $0.981^{+0.150}_{-0.122}$ & $1.051^{+1.087}_{-0.488}$\\
&&&\\
\hline
\end{tabular}

{\small Note -- Masses are in $h^{-1} M_{\sun}$.}
\end{table}

In Table \ref{tab:popsims_results}, the derived best-fit parameters are listed along with their unmarginalized 68\% errors, constraining $f$ and $f_{\rm cross}$ to within $10-15\%$.  The table shows that the true values of all the parameters fall within the 68\% confidence intervals.  Deviations from true parameter values may be attributed in part to degeneracies between parameters, to sample variance, or to approximations in the calculation of the correlation function, such as use of spherical halo exclusion in place of elliptical exclusion. 

\begin{figure*}
\centering
\resizebox{3.4in}{!}
	{\includegraphics{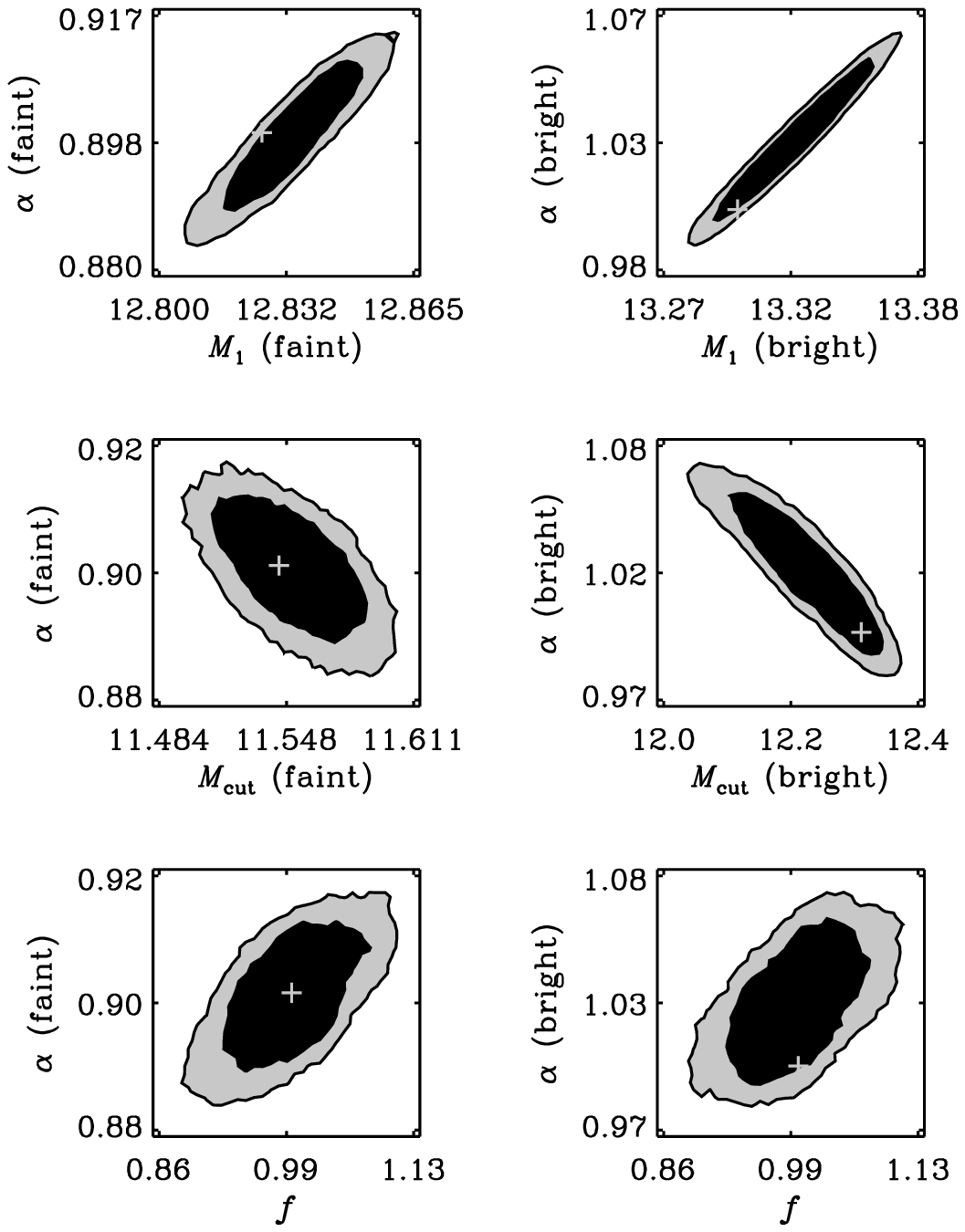}}
\resizebox{3in}{!}{\includegraphics{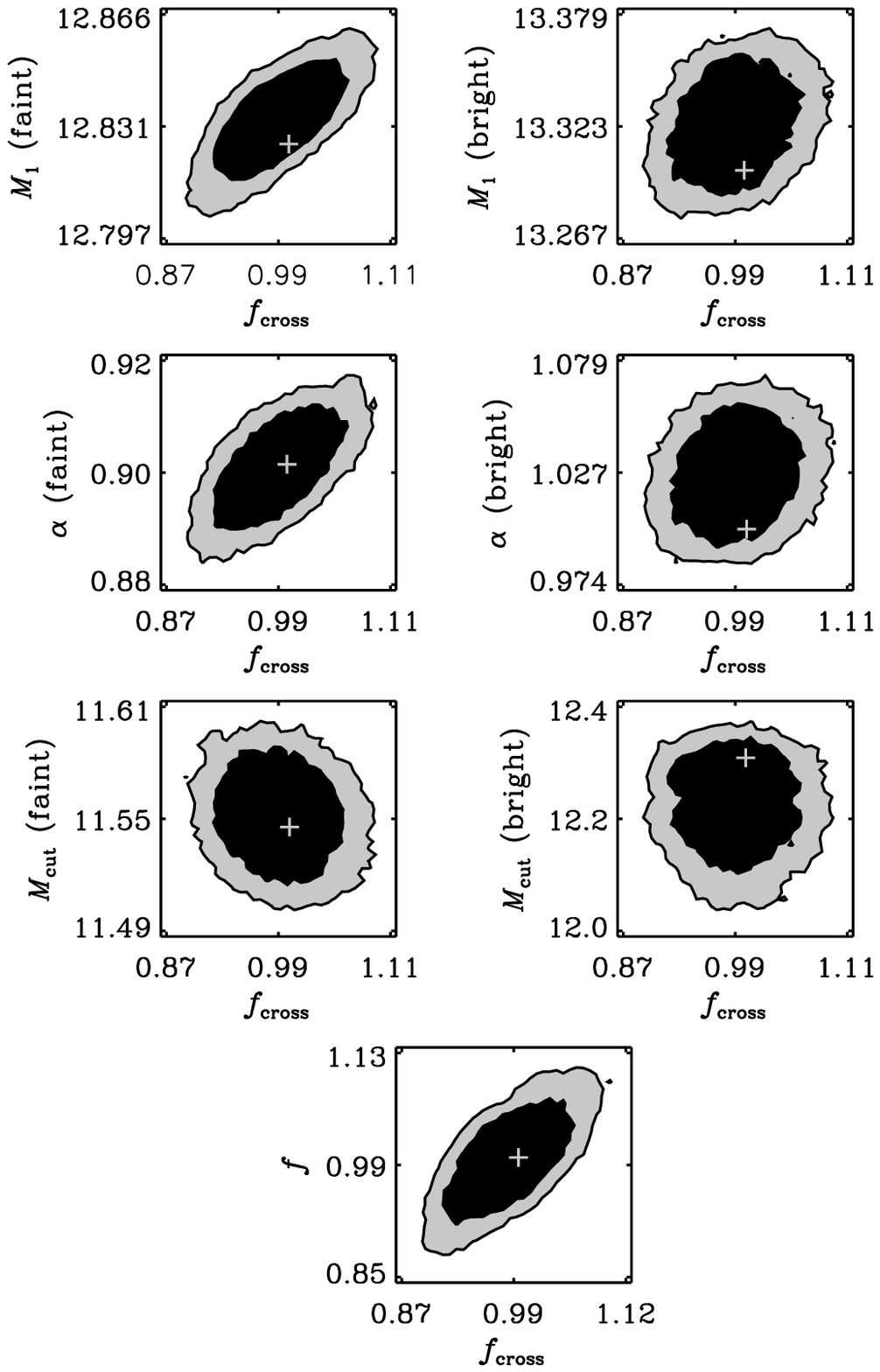}}
\caption{The 68 and 90\% marginalized confidence intervals are shown for pairs of parameters in the idealized data set.    All masses are shown in log($h^{-1} M_{\sun}$).  Crosses indicate the input parameter values.   }
\label{fig:400.best_fit.deg}
\end{figure*}

Degeneracies between pairs of parameters are investigated in Fig. \ref{fig:400.best_fit.deg}.  In each panel, the 68\% and 90\% confidence intervals are shown, marginalizing over all other parameters.  $f$ and $f_{\rm cross}$ are more degenerate with the faint HOD parameters than the bright parameters.  The faint sample has many more objects than the bright sample, so its constraining ability is greater.  In addition,  $f_{\rm cross}$  is only constrained by the one-halo, central-satellite term of the cross-correlation, which depends only on the faint HOD parameters.  However, the degeneracies between HOD parameters -- $M_{1}$, $\alpha$, $M_{\rm cut}$ -- in a single sample are much larger.  Finally, the degeneracy between $f$ and $f_{\rm cross}$ is pronounced and behaves such that larger values of $f$ are accompanied by larger values of $f_{\rm cross}$ and vice versa.  

The behavior of these degeneracies can be seen from the form of the HOD itself.  Compared to an arbitrary satellite function, $\langle N_{\rm sat} \rangle$, increasing the parameter $M_{1}$ reduces the overall number of satellites.  This, however, can be compensated at masses larger than $M_{1}$ by increasing the slope of the function $\alpha$.  At smaller masses, increasing $M_{1}$ and $\alpha$ reduces the fraction of central-satellite pairs.  Decreasing the roll-off parameter, $M_{\rm cut}$, marginally increases the fraction of central-satellite pairs.  Thus, larger $M_{1}$ indicates larger $\alpha$ and smaller $M_{\rm cut}$.  In relation to the parameter $f$, increasing $\alpha$ tilts the correlation function to pairs in halos with larger masses and smaller concentrations;  increasing $f$ and $f_{\rm cross}$ counteracts this by boosting the concentrations. 

\subsection{Tests Using SDSS-Sized Mock Data Set}

A subsample of the simulation box that is 115 $h^{-1}$ Mpc on a side is equal in volume to the SDSS data sample.  Using 196 jackknife samples (similar to the 205 samples used in the data) and estimating all correlations and covariances from the simulation, the size of the error bars are larger than in the case of the idealized box.

The results of the parameter estimation are shown in Table \ref{tab:popsims_results} and are particularly striking in the case of the roll-off parameter, $M_{\rm cut}$, where the parameter space of likely values extends to $M_{\rm cut} < M_{\rm min}$.  Once $M_{\rm cut}$ drops below the minimum mass for central galaxies, it no longer has any effect on the correlation function, resulting in extremely large errors on the parameter estimation.  At the volume of the SDSS data set, $f=1$ can be distinguished from $f=2$, but values of $f_{\rm cross}=0.5-2$ are consistent.   

The resulting errors in the parameter estimation are not directly applicable to the SDSS data.  In addition to the issue of sample variance, the populated simulations are an idealized dataset where the halos are exactly populated according to a HOD, there is no variance in the concentration of halos, and the cosmology of the populated simulation is the same as that used in the model. 
The presented analysis, however, demonstrates approximately the degree to which the HOD and the radial distribution of satellites can be constrained with the current volume-limited SDSS sample.

\subsection{Tests Including Systematics and Varying $f$ and $f_{\rm cross}$}
\label{sec:systematictest}

Using the SDSS-sized mock data set, I include the effects of the systematics in the SDSS data:  1.) issues with the sky subtraction (see Section \ref{sec:corr}) and 2.) incompleteness in the spectroscopic catalog (see Section \ref{sec:sdss}).  

As discussed previously, the sky subtraction problem is angle dependent, over-predicting pairs at separations less than $20-30\arcsec$ and under-predicting by $\sim5\%$ pairs at separations between $40-90\arcsec$.  At the median redshift of the data, $30\arcsec$ is $16 h^{-1}$ kpc.  The overestimate in pairs at small angular separations is unconstrained, but in the data $\sim25\%$ of the objects in the first bin of the correlation function (from 10 to $16h^{-1}$ kpc) are found at separations less than $20\arcsec$.  I assume that all of the 25\% are spurious pairs and add that fraction of pairs into the mock data at separations less than $16 h^{-1}$ kpc.  $40-90\arcsec$ is $22-50 h^{-1}$ kpc at the median redshift of the data.  In the mock data, I remove 5\% of the pairs with separations between 22 and $50h^{-1}$ kpc.

In addition to including these refinements, I test different values of $f$ and $f_{\rm cross}$ in order to show that a range of values can be reproduced.  The catalogs produced are  MA1 ($f=0.5$, $f_{\rm cross}=1$);  MB1 ($f=1$, $f_{\rm cross}=1.5$);  and MC1 ($f=1$, $f_{\rm cross}=0.5$).  I also create a sample MA2 where the global completeness is exactly 90\%.  This overstates the effect of incompleteness in the data where I use all areas with completeness better than 90\%. 

Results are shown in Table \ref{tab:popsims_results2}.  While the errors in these cases can be larger than the errors without accounting for systematic effects, there are no biases in the estimated HOD values and a full range of values for $f$ and $f_{\rm cross}$ are recoverable from the data.

\begin{table}[h]
\caption{Unmarginalized, Best-Fit $f$ and $f_{\rm cross}$ for Mock Data Sets}
\label{tab:popsims_results2}
\begin{tabular}{lllll}
\hline
\multicolumn{1}{c}{Mock Sample} &
\multicolumn{2}{c}{$f$} &
\multicolumn{2}{c}{$f_{\rm cross}$}\\
\multicolumn{1}{c}{}&
\multicolumn{1}{c}{Input}&
\multicolumn{1}{c}{Best-Fit}&
\multicolumn{1}{c}{Input} &
\multicolumn{1}{c}{Best-Fit} \\
\hline\hline
&&&\\
MA1 & 0.5 & $0.317^{+0.314}_{-0.179}$ & 1 & $1.021^{+1.494}_{-0.387}$\\
&&&& \\
MB1 & 1 & $0.852^{+0.609}_{-0.333}$ & 1.5 & $2.166^{+5.372}_{-0.946}$\\
&&&&\\
MC1 & 1 & $0.987^{+0.613}_{-0.473}$& 0.5 & $0.813^{+0.960}_{-0.420}$ \\
&&&&\\
MA2 & 0.5 & $0.367^{+0.257}_{-0.233}$ & 1 & $1.091^{+1.219}_{-0.477}$\\
&&&&\\
\hline
\end{tabular}

{\small Note -- Errors are 68\% confidence intervals.}
\end{table}

\section{Results}
\label{sec:results}


The analysis applied to the test samples can be repeated for the SDSS samples.  First, the values of  $M_{\rm max(cen)}$ for the bright and faint sample must be estimated.  With those values, the fiducial results for the HOD parameters can be found.  Then, the robustness of the best-fit values of $f$ and $f_{\rm cross}$ is tested.  Finally, the results are compared to weak lensing results.

\subsection{Estimating $M_{\rm max(cen)}$}

The maximum halo mass, $M_{\rm max(cen)}$, for the bright and faint samples must be measured from additional SDSS samples as described in Section \ref{sec:hm}.  A luminosity-threshold sample of $M_{r} < -21$ is used to calculate $M_{\rm max(cen)}$ for the bright sample and a luminosity-threshold sample of $M_{r} < -19$ is used for the faint sample.  For the $M_{r} < -19$ sample, I fit 4 parameters, $M_{1}$, $\alpha$, $M_{\rm cut}$, and $f$.  For the $M_{r} < -21$ sample, the number of objects with projected separations less than $r_p = 0.1 h^{-1}$ Mpc is small, so constraints on $f$ are extremely difficult to obtain, and I set $f=1$ and fit only 3 parameters.  

\begin{table}[h]
\caption{HOD Parameters for SDSS Luminosity-Threshold Samples}
\label{tab:mmax_test}
\begin{tabular}{lll}
\hline
Parameter &
$M_{r} < -21$ &
$M_{r} < -19$ \\
\hline
$M_{\rm max(cen)}$& $1.03 \times 10^{13}$  & $5.44 \times 10^{11}$ \\
$M_{1}$ & $1.93 \times 10^{14}$  &$6.09 \times 10^{12}$ \\
$\alpha$ & 1.25 &  0.97 \\
$M_{\rm cut}$ & $7.24 \times 10^{7}$ & $1.78\times 10^{9}$\\
$f$ & ----- &  0.40 \\
68\% $M_{\rm max(cen)}$ minimum &  $9 .45\times 10^{12}$  & $3.92 \times 10^{11}$ \\
68\% $M_{\rm max(cen)}$ maximum &  $1.13\times10^{13}$ & $6.33\times 10^{11}$ \\
\hline
\end{tabular}

{\small Note -- All masses are in $h^{-1} M_{\sun}$.}
\end{table}

The best-fit values and the smallest and largest values for $M_{\rm max(cen)}$ for the unmarginalized parameter space that contains 68\% of the values estimated by MCMC are listed in Table \ref{tab:mmax_test}.  Even using luminosity-threshold samples, however, the range of values for the bright and faint $M_{\rm max(cen)}$ is rather large, with some $\sim$20-30\% variation in the faint sample and less than $\sim 10\%$ variation in the bright sample.   Previously published results using different volumes of SDSS data find similar, but not identical results.  For example, \citet{zehavi_etal05} -- which fixes $f=1$ -- find for their $M_r < -21$ sample that $M_{\rm max(cen)} = 5.25\times10^{12} h^{-1} M_{\sun}$, $M_{1} = 1.23\times10^{14} h^{-1} M_{\sun}$, and $\alpha=1.39$.  For their $M_r < -19$ sample, they find $M_{\rm max(cen)} = 3.89\times10^{11} h^{-1} M_{\sun}$, $M_{1} = 8.7\times10^{12} h^{-1} M_{\sun}$, and $\alpha=1.08$.

\subsection{Fiducial Results for $f$ and $f_{\rm cross}$}

\begin{table}
\caption{Unmarginalized, Best-Fit Parameters for SDSS Data Using Best-Fit $M_{\rm max(cen)}$}
\label{tab:data_results}
\begin{tabular}{lll}
\hline
\multicolumn{1}{c}{Parameter} &
\multicolumn{2}{c}{Best-Fit with 68\% errors}\\
\multicolumn{1}{c}{}&
\multicolumn{1}{c}{all bins} &
\multicolumn{1}{c}{omitting first bin} \\
\hline\hline
faint: & &\\
~~~~~~~~~~log($M_{1}$) & $12.75^{+0.14}_{-0.09}$ & $12.75^{+0.17}_{-0.10}$\\
&&\\
~~~~~~~~~~$\alpha$        & $0.90^{+0.07}_{-0.06}$ & $0.90^{+0.08}_{-0.06}$\\
&&\\
~~~~~~~~~~log($M_{\rm cut}$)  & $9.09^{+2.16}_{-0.09}$ &$9.48^{+1.76}_{-0.48}$\\
&&\\
\hline
bright: & &\\
~~~~~~~~~~log($M_{1}$) & $13.29^{+0.12}_{-0.18}$& $13.29^{+0.13}_{-0.26}$ \\
&&\\
~~~~~~~~~~$\alpha$       & $0.96^{+0.12}_{-0.13}$ &$0.99^{+0.12}_{-0.26}$\\
&&\\
~~~~~~~~~~log($M_{\rm cut}$)  & $10.21^{+2.19}_{-1.21} $ &$10.996^{+1.57}_{-1.96}$\\
&&\\
\hline
&&\\
$f$                 & $0.41^{+0.53}_{-0.17} $ & $0.44^{+0.51}_{-0.24}$\\
&&\\
$f_{\rm cross}$     & $0.95^{+1.20}_{-0.41}$ &$0.83^{+3.18}_{-0.42}$\\
& &\\
\hline
\end{tabular}

{\small Note -- Masses are in $h^{-1} M_{\sun}$.}
\end{table}

\begin{figure}
\centering
\resizebox{3.5in}{!}
	{\includegraphics{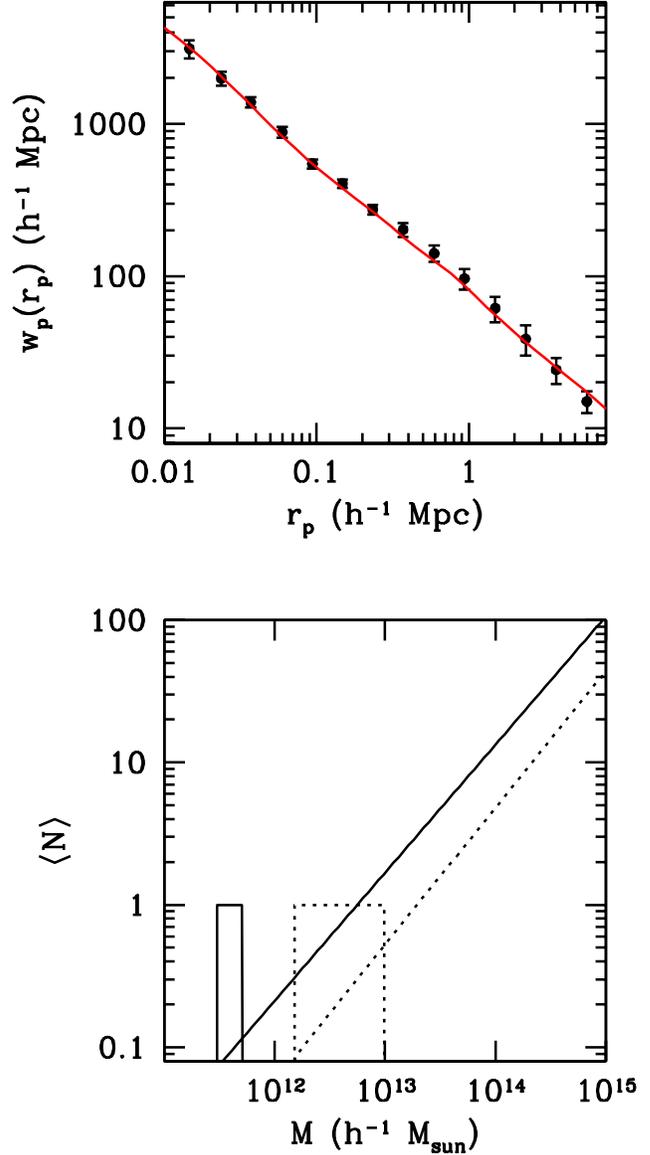}}
\caption{{\it Top:}  The estimated cross-correlation ({\it circles}) compared to the halo model cross-correlation using the best-fit HOD parameters ({\it solid line}).  {\it Bottom:}  The shape of the HOD for the bright sample ({\it dotted}) and the faint sample ({\it solid}).  }\label{fig:hod}
\end{figure}

\begin{figure}
\centering
\resizebox{3.5in}{!}
	{\includegraphics{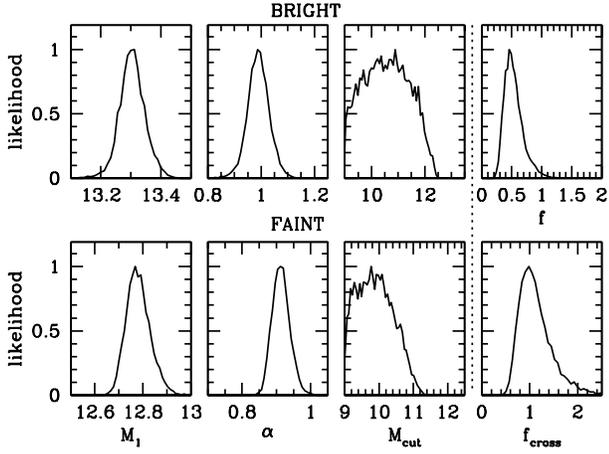}}
\caption{The normalized, marginalized likelihoods for all parameters ($M_{1}$, $\alpha$, and $M_{\rm cut}$ for the bright and faint samples and $f$ and $f_{\rm cross}$) in the SDSS data.  All masses are shown in log($h^{-1} M_{\sun}$). }\label{fig:marg}
\end{figure}

Using the best-fit values for $M_{\rm max(cen)}$, the best-fit HOD parameters for the luminosity-binned samples are found and shown in Figure \ref{fig:hod} along with the best-fit cross-correlation.  The unmarginalized best-fit parameters are listed in Table \ref{tab:data_results}.  The marginalized likelihoods for the parameters are shown in Figure \ref{fig:marg}.  

As expected from the tests of the mock data sets, the value of $f_{\rm cross}$ is not well-constrained by the data, although the best-fit values suggest that the galaxy distribution is consistent with the expected DM distribution, $f_{\rm cross}=1$.  There is a wide range of values from $\sim .5$ to $\sim 2$ within the unmarginalized 68\% error ellipse.  Extremely small values, however, are not supported ($f_{\rm cross} < 0.5$).  $f$, as expected, is better constrained, with values from $\sim .2$ and $\sim 1$ within the unmarginalized 68\% error ellipse.  $f<1$ is very probable.  This may suggest that the distribution of galaxies is shallower than the dark matter distribution, which would still be consistent with previous work at galaxy and cluster scales \citep[e.g.,][]{nagai_kravtsov05,chen_etal06}.  Finally, recalling the degeneracy between $f$ and $f_{\rm cross}$, if one were to guess that the best-fit estimate of $f$ was biased to smaller values, one would also have to assume that the most likely value of $f_{\rm cross}$ is also too small.

\subsection{Tests of the Robustness of the Fiducial Results} \label{sec:robustnessresults}

In addition to the fiducial data set, I test a shortened data set, omitting the first bin in the correlation functions (physical separations smaller than 16 $h^{-1}$ kpc), with unmarginalized results also shown in Table \ref{tab:data_results}.  As discussed in Section \ref{sec:corr}, this bin is most affected by sky subtraction issues in the photometry.  Omitting it does not appreciably change any of the results, aside from inflating the error bar on $f_{\rm cross}$, the parameter most sensitive to small projected separations.  

Because $M_{\rm max(cen)}$ is estimated from different luminosity samples than the faint and bright samples, I test the effects of the variances in $M_{\rm max(cen)}$ estimates.  I run the MCMC for four samples, changing in turn the bright and faint $M_{\rm max(cen)}$ to the lower and the upper value in the 68\% unmarginalized confidence intervals and leaving the other $M_{\rm max(cen)}$ fixed to the fiducial value.  The test samples are described and results are presented in Table \ref{tab:data_resultsA_D}.  Here, the parameter estimates look very similar to the fiducial values, except in the case where a lower estimate of $M_{\rm max(cen)}$ in the bright sample was used.  In all cases, however, the errors overlap and all show consistent results of $f_{\rm cross}$.  

\begin{table}
\caption{Unmarginalized, Best-Fit Parameters for SDSS Data Varying $M_{\rm max(cen)}$}
\label{tab:data_resultsA_D}
\begin{tabular}{lllll}
\hline
Sample&
A $^{a}$ &
B $^{b}$ &
C $^{c}$ &
D $^{d}$\\
\hline\hline
faint: &&&&\\
~~~~~~~~~~log($M_{1}$)& $12.75^{+0.16}_{-0.08}$ & $12.95^{+0.19}_{-0.15}$  & $12.76^{+0.13}_{-0.10}$ & $12.75^{+0.15}_{-0.11}$ \\
&&&&\\
~~~~~~~~~~$\alpha$        & $0.90^{+0.08}_{-0.05}$ & $1.00^{+0.12}_{-0.10}$ & $0.91^{+0.06}_{-0.06}$ & $0.90^{+0.08}_{-0.06}$ \\
&&&&\\
~~~~~~~~~~log($M_{\rm cut}$)  & $9.50^{+1.69}_{-0.50}$& $9.01^{+2.05}_{-0.01}$& $9.94^{+1.34}_{-0.94}$& $9.34^{+1.69}_{-0.34}$ \\
&&&&\\
\hline
bright: &&&& \\
~~~~~~~~~~log($M_{1}$) & $13.29^{+0.13}_{-0.18}$ & $13.52^{+0.14}_{-0.15}$     & $13.29^{+0.12}_{-0.21}$ & $13.31^{+0.11}_{-0.16}$ \\
&&&&\\
~~~~~~~~~~$\alpha$       & $0.97^{+0.13}_{-0.14}$& $1.17^{+0.17}_{-0.14}$   & $0.97^{+0.11}_{-0.17}$& $1.00^{+0.10}_{-0.14}$ \\
&&&&\\
~~~~~~~~~~log($M_{\rm cut}$)  & $11.34^{+1.06}_{-2.34} $ & $9.70^{+2.94}_{-0.70} $ & $10.71^{+1.81}_{-1.71} $ & $9.94^{+2.51}_{-0.94} $\\
&&&&\\
\hline
&&&&\\
$f$                 & $0.41^{+0.58}_{-0.17} $& $0.92^{+0.89}_{-0.50} $              & $0.43^{+0.47}_{-0.20} $& $0.47^{+0.55}_{-0.23} $\\
&&&&\\
$f_{\rm cross}$     & $0.89^{+1.20}_{-0.39}$& $0.99^{+1.87}_{-0.51}$       & $1.00^{+1.28}_{-0.49}$ & $0.88^{+1.19}_{-0.39}$\\
& &&&\\
\hline
\end{tabular}

{\small Note -- All masses are in $h^{-1} M_{\sun}$.\\
$^a$ For the bright sample, $M_{\rm max(cen)}=1.13\times 10^{13} h^{-1} M_{\sun}$.\\
$^b$ For the bright sample, $M_{\rm max(cen)}=9.45\times 10^{12} h^{-1} M_{\sun}$.\\
$^c$ For the faint sample, $M_{\rm max(cen)}=6.33\times 10^{11} h^{-1} M_{\sun}$.\\
$^d$ For the faint sample, $M_{\rm max(cen)}=3.92\times 10^{11} h^{-1} M_{\sun}$.}
\end{table}

In Figure \ref{fig:marg}, the parameter $M_{\rm cut}$ has an oddly shaped and broad likelihood distribution (discussed briefly in the previous section).  The marginalized likelihoods for $M_{\rm cut}$ in the SDSS data suggest rather small values.  As noted previously, when $M_{\rm cut} < M_{\rm min}$, $M_{\rm cut}$ contributes nothing to the HOD.  One may be tempted to suggest that the small $M_{\rm cut}$ values are due to poor constraints of the data.  However, small values may be well-motivated, at least for the bright sample.  As the bright $M_{\rm cut}$ increases, the amplitude at small scales in the autocorrelation, where the one-halo central-satellite term dominates, decreases.  The amplitude in the cross-correlation, however, is unaffected, since the central-satellite term in the cross-correlation samples the part of the faint sample satellite distribution that is unaffected by $M_{\rm cut}$.  Thus, large values of $M_{\rm cut}$ for the bright sample would cause the bright autocorrelation and the cross-correlation to intersect, which is not supported by the data (see Fig. \ref{fig:data}).

\subsection{Comparison to Weak Lensing Results}
\label{sec:weaklens}

If satellites are found around galaxies with the same distribution as the dark matter profile, the cross-correlation of bright and faint galaxies may be related to the galaxy-mass cross-correlation.  The projected galaxy-mass cross-correlation is proportional to the weak lensing measurement of the surface density contrast, $\Delta \Sigma(r_p)$.  Given the larger errors in the weak lensing measurements at the scales at which the satellite distribution and the dark matter distribution can be compared, a comparison of simple power-law fits may be sufficient, assuming a pure power-law density profile for $\Sigma(r_p)$ at and below those scales.  

In comparison to the weak lensing measurement, the cross-correlation of bright and faint galaxies has a somewhat steeper overall power-law slope than the weak-lensing measuring.  For example, \citet{sheldon_etal04} find a power-law slope of $\gamma=-0.76 \pm 0.05$ for projected radii between 0.025 and 10 $h^{-1}$ Mpc and their lens sample is peaked near an $r$-band magnitude of $-21$.  In addition, comparisons to the measurements from \citet{mandelbaum_etal06} for a lens sample with $r$-band magnitudes between $-21$ and $-20$ show similar power-law slopes, $\gamma=-0.79 \pm 0.04$ for projected radii between 0.02 and 2 $h^{-1}$ Mpc.  Both measurements are shallower than the power-law slope of the cross-correlation of bright and faint galaxies, which show $\gamma=-0.86 \pm 0.02$ at projected radii between 0.01 and 16 $h^{-1}$ Mpc.\footnote{The power-law slope $\gamma$ for the cross-correlation is calculated by fitting for $w_p(r_p) = A \times r_p^{\gamma}$. } 

The cross-correlation function, however, is more sensitive to the distribution of satellites around central galaxies at smaller separations.  The best-fit power-law slope for the projected cross-correlation remains fairly consistent, with $\gamma=-0.88 \pm 0.04$ for projected radii between 0.01 and 0.25 $h^{-1}$ Mpc and $\gamma=-0.93 \pm 0.07$ for projected radii between 0.01 and 0.1$h^{-1}$ Mpc.  \citet{mandelbaum_etal06} shows larger error bars at small radii, with rough consistency with my results, $\gamma =-0.98 \pm 0.10$ for projected radii between 0.02 and 0.2 $h^{-1}$ Mpc and $\gamma =-0.77 \pm 0.20$ for projected radii between 0.02 and 0.1 $h^{-1}$Mpc.  The best-fit power-laws for the cross-correlation of bright and faint galaxies and the galaxy-mass cross-correlation around bright galaxies are statistically consistent at the most significant separations.  This is consistent with a result of $f_{\rm cross} =1$ -- the satellite distribution follows the DM halo distribution.

\section{Conclusions}
\label{sec:conclusions}

The spatial distribution of satellite galaxies around $\sim L_*$  galaxies can illuminate the relationship between satellite galaxies and dark matter subhalos and aid in developing and testing galaxy formation models.  The projected cross-correlation of bright galaxies with faint galaxies offers a promising avenue to putting constraints on the radial distribution of satellite galaxies.  

Results from SDSS data of 5104 ${\rm deg^2}$ on the sky suggest that the spatial distribution of galaxies, parameterized by $f$ and $f_{\rm cross}$, is generally consistent with the dark matter distribution for halos in a flat $\Lambda$CDM cosmology with $\Omega_{\rm m}=0.3$, $\Omega_{\rm b}=0.04$, $\sigma_{8}=0.9$, 
and $h=0.7$.  $f_{\rm cross}$ parameterizes the distribution of `faint' satellites galaxies around `bright' central galaxies.  There are no strong constraints on the value of $f_{\rm cross}$, however, with a wide range of values from $\sim .5$ to $\sim 2$ in the unmarginalized 68\% error ellipse, where $f_{\rm cross} =1$ would indicate a distribution equivalent to the DM halo profile.  $f$, on the other hand, parameterizes the distribution of both bright and faint satellite galaxies around central galaxies of all sizes.  Constraints on the value of $f$ are significantly better -- ruling out $f=2$ -- and are suggestive of somewhat smaller concentrations for galaxy distributions compared to the concentration of dark matter halos ($f<1$).  The spatial distribution of galactic satellites remains poorly constrained and future studies in this field will require a greater volume of data.  

The results presented here use a flat  $\Lambda$CDM cosmology with $\sigma_8=0.9$ which is consistent with the results of the Wilkinson Microwave Anisotropy Probe (WMAP) 1-Year Data \citep{spergel_etal03}.  Recent results from the WMAP 5-Year Data supports a smaller $\sigma_8=0.8$ \citep{dunkley_etal08}.  \citet{duffy_etal08} claim that halo concentrations using a WMAP5 cosmology are lower than in a WMAP1 cosmology by 16-23\%.  If halos are less concentrated than the model predicts, the fit can compensate by decreasing $f$.  Since $f$ is defined relative to the profile of the DM halo, this implies that the true $f$ and $f_{\rm cross}$ are actually larger than the estimated $f$ and $f_{\rm cross}$.  The difference in halo concentration is, however, fairly modest compared to the error estimate in $f$ and $f_{\rm cross}$.  

The results of this study have implications for previous works on constraining the satellite spatial distribution, galaxy clustering models, and weak lensing studies of the galaxy-mass cross-correlation.  For example, the results are consistent with those of \citet{chen_etal06}:  satellite galaxies are distributed in a way that is consistent with or less concentrated than the dark matter distribution.  The results of \citet{chen_etal06} apply only to isolated galaxies, while the cross-correlation approach applies to any parent galaxy.  This suggests that the environment of parent galaxies is relatively unimportant.  My result is also generally consistent with cluster-sized simulations which include cooling and star formation.  In such simulations, while the distribution of dark matter subhalos is less concentrated than the dark matter, the stellar components of satellite galaxies more closely follow the dark matter distribution \citep{nagai_kravtsov05,maccio_etal06}.  However, the possible discrepancy in values for $f$ and $f_{\rm cross}$ suggest that the spatial distribution of a class of satellite objects may depend on the size of the host halo:  galaxy-sized halos, groups and clusters.

\citet{chen08} suggest that the projected radial distribution of satellite galaxies depends upon satellite color, such that redder satellites have a significantly steeper density profile than bluer satellites.  A cross-correlation analysis with a different selection function for satellites then should show such a color dependence.  In order to achieve better constraints on the spatial distribution of satellite galaxies and its environmental and color dependency, larger data sets are required.  One possibility for increasing the volume of data is to apply a similar analysis to the photometric redshift catalog, instead of limiting the sample to objects with spectroscopic redshifts.  The SDSS photometric redshift catalog goes far deeper than the spectroscopic catalog and without the spectroscopic catalog's fiber collision problem.  However, photometric redshifts are less accurate and may complicate the statistical error analysis.  Given that future large surveys such as the Dark Energy Survey and the Panoramic Survey Telescope \& Rapid Response System (Pan-STARRS) which will produce significantly more photometric redshifts than currently available, it would be interesting to perform a cross-correlation analysis with photometric redshifts.


\begin{acknowledgements}
An incredible number of people need to acknowledged for their 
invaluable contributions.  I would like to thank Jeremy Tinker for allowing me to use and modify his halo model code, and 
Zheng Zheng for his help with additions to the analysis code.  I would also like to thank Michael Warren 
for allowing me to use his simulations, Risa Wechsler and Charlie 
Conroy for providing me with luminosity assigned galaxy catalogs, and  
Michael Blanton and Morad Masjedi for useful discussions about the data. 
In addition, Erin Sheldon and Rachel Mandelbaum were extremely helpful 
in discussions of the data, its systematics, and weak lensing observations,
as was Ryan Scranton in sharing his code to create jackknife samples. Eduardo Rozo, 
Douglas Rudd, and 
Sean Andrews each provided useful advice on a variety of topics. 
Finally, I would like to thank my adviser, Andrey Kravtsov, for 
his guidance and insight.  

This research was carried out at the University of Chicago, Kavli
Institute for Cosmological Physics and was supported (in part) by
grant NSF PHY-0114422. KICP is an NSF Physics Frontier Center. JC 
was supported by the National Science Foundation (NSF) under
grants No.  AST-0206216, AST-0239759 and AST-0507666, and by NASA
through grant NAG5-13274. 

Funding for the Sloan Digital Sky Survey (SDSS) has been provided by
the Alfred P. Sloan Foundation, the Participating Institutions, the
National Aeronautics and Space Administration, the National Science
Foundation, the U.S. Department of Energy, the Japanese
Monbukagakusho, and the Max Planck Society. The SDSS Web site is
http://www.sdss.org/.

The SDSS is managed by the Astrophysical Research Consortium (ARC) for
the Participating Institutions. The Participating Institutions are The
University of Chicago, Fermilab, the Institute for Advanced Study, the
Japan Participation Group, The Johns Hopkins University, Los Alamos
National Laboratory, the Max-Planck-Institute for Astronomy (MPIA),
the Max-Planck-Institute for Astrophysics (MPA), New Mexico State
University, University of Pittsburgh, Princeton University, the United
States Naval Observatory, and the University of Washington.
\end{acknowledgements}

\bibliography{draft}

\end{document}